\documentclass[hyper]{JHEP} 
\usepackage{amsmath,amssymb,amsfonts}

\usepackage{epsfig}

\newcommand\fverb{\setbox\pippobox=\hbox\bgroup\verb}

\newcommand\fverbdo{\egroup\medskip\noindent%

            \fbox{\unhbox\pippobox}\ }

\newcommand\fverbit{\egroup\item[\fbox{\unhbox\pippobox}]}

\newbox\pippobox

\title{Note About Consistent Extension of
  Quasidilaton Massive Gravity}
\author{J. Kluso\v{n}\\
Department of
Theoretical Physics and Astrophysics\\
Faculty of Science, Masaryk University\\
Kotl\'{a}\v{r}sk\'{a} 2, 611 37, Brno\\
Czech Republic\\
E-mail: \email{klu@physics.muni.cz}}
\preprint{}

 \abstract{This note is devoted to the Hamiltonian analysis of
  extension of  quasidilaton massive gravity as was proposed recently in
[arXiv:1306.5502]. We show that for given formulation of the theory
the additional primary constraint that is responsible for the
elimination of the Boulware-Deser ghost is missing. We compare this
situation with quasidilaton massive gravity. Finally we  propose
ghost free extension of quasidilaton massive gravity.}
\keywords{Massive Gravity, \ Hamiltonian Formalism}

\def\tf{\tilde{f}}

\def\mC{\mathcal{C}}
\def\be{\begin{equation}}

\def\ee{\end{equation}}

\def\tn{\tilde{n}}
\def\tx{\tilde{x}}
\def\tK{\tilde{K}}
\def\mR{\mathcal{R}}
\def\mM{\mathcal{M}}
\def\tSigma{\tilde{\Sigma}}
\def\bea{\begin{eqnarray}}

\def\eea{\end{eqnarray}}

\def\mH{\mathcal{H}}

\def\bx{\mathbf{x}}
\def\by{\mathbf{y}}

\newcommand{\hg}{\hat{g}}

\newcommand{\tD}{\tilde{D}}
\newcommand{\mG}{\mathcal{G}}

\def\pb #1{\left\{#1\right\}}

\begin{document}
\section{Introduction and Summary}\label{first}
Recently new version of the full non-linear massive gravity that was
found by de Rham, Gabadadze and Toley (dRGT)
\cite{deRham:2010ik,deRham:2010kj} provides the positive answer to
the question whether graviton can have a non-zero mass. In fact,
among many remarkable properties there is the crucial one which is
the absence of the Boulware-Deser ghost
\cite{Boulware:1973my,Boulware:1972zf}.

The consistent massive gravity could also provide a possible
explanation of the observed acceleration of the cosmic expansion
which is one of the greatest mysteries in modern cosmology. It is
tempting to speculate that the finite graviton mass could be a
source of the accelerated expansion of the universe. For that reason
it is great interest to formulate theoretically consistent
cosmological scenario in massive gravity that is also in agreement
with the observations. Unfortunately it was recently shown that all
homogeneous and isotropic cosmological solutions in dRGT theory are
unstable \cite{DeFelice:2012mx}, see also
\cite{Koyama:2011wx,Tasinato:2012ze,Khosravi:2013axa}.

In order to resolve this problem we have two possible options:
Either to break homogeneity \cite{D'Amico:2011jj} or isotropy
\cite{Gumrukcuoglu:2012aa,DeFelice:2013awa} or to extend the theory
as in \cite{D'Amico:2012zv,Huang:2012pe}. Recently in
\cite{DeFelice:2013tsa} A. De Felice and S. Mukohyama proposed new
extension of quasidilaton massive gravity
 that could provide stable and
self-accelerating homogeneous and isotropic cosmological solution.
They further  argued that given extension belongs to the class of
models studied in \cite{Gabadadze:2012tr} that are free from the
Boulware-Deser ghosts. However the explicit Hamiltonian analysis of
given theory was not performed in \cite{DeFelice:2013tsa}.

The goal of this paper is to reconsider the problem of the
Boulware-Deser ghost in the model \cite{DeFelice:2013tsa}.
 We present an evidence that for the action that was
   introduced in \cite{DeFelice:2013tsa}
 the Boulware-Deser ghost cannot be eliminated.
 More precisely, performing the Hamiltonian analysis of
this model with time dependent quasidilaton we find that primary
constraint, that is responsible for the elimination of the ghost in
 St\"{u}ckelberg formulation of non-linear massive gravity
\cite{Kluson:2012zz,Kluson:2012wf,Hassan:2012qv,Kluson:2013hoa}  is
missing \footnote{For related  work, see \cite{Huang:2013mha}.}.
This result implies that generally Boulware-Deser ghost is present.
On the other hand we show that this additional constraint emerges
when $\omega\neq 0$ and when $\alpha_\sigma=0$ that corresponds to
the quasidilaton massive gravity. We also propose model of
consistent extension of  the quasidilaton non-linear massive gravity
that can be considered as the generalization of the coupling between
massive gravity and the galileon \cite{Andrews:2013ora} and we argue
that given theory is ghost free, following \cite{Kluson:2013hoa}.

This paper is organized as follows. In the next section
(\ref{second}) we review the basic facts about extension of
quasidilaton non-linear massive gravity as was proposed in
\cite{DeFelice:2013tsa}. Then we proceed to the Hamiltonian analysis
of given theory and argue that there is no scalar primary constraint
that could eliminate the Boulware-Deser ghost. In section
(\ref{third}) we perform the Hamiltonian analysis of quasidilaton
non-linear massive gravity when  we find that in this case
 there is an additional primary constraint. This result shows  that
the quasidilaton massive theory as was proposed in
\cite{D'Amico:2012zv} is ghost free at least in  their minimal
version. Finally in section (\ref{fourth}) we propose the extension
of the quasidilaton massive gravity that is ghost free and that can
be considered as the generalization of the proposal
\cite{DeFelice:2013tsa}. It would be extremely interesting to
analyze cosmological consequences of this theory.

\section{ Extension of Quasidilaton Massive Gravity}\label{second}
In this section we review basic facts about extension of
quasidilaton massive gravity as was proposed in
\cite{DeFelice:2013tsa}. For simplicity we restrict ourselves to the
minimal form of the massive gravity keeping in mind that its
generalization is straightforward.

 Explicitly, let us consider following
action
\begin{eqnarray}\label{Smass}
S&=&S_{m.g.}+S_{\sigma} \ , \nonumber \\
 S_{m.g.}&=& M_p^2 \int d^4x\sqrt{-\hg}
[{}^{(4)}R+2m^2(3-\Omega(\Phi)\sqrt{\hg^{-1}\tf})] \ , \nonumber \\
\end{eqnarray}
where  $\tf_{\mu\nu}$ was introduced in
\cite{DeFelice:2013tsa}
\begin{equation}
\tf_{\mu\nu}=f_{\mu\nu}-\frac{\alpha_\sigma}{M_p^2 m^2}
e^{-2\sigma/M_p}\partial_\mu\sigma\partial_\nu\sigma \ , \quad
f_{\mu\nu}=\partial_\mu\phi^a\partial_\nu\phi^b\eta_{ab} \ ,
\end{equation}
where $\phi^a, a,b=0,1,2,3$ are St\"{u}ckelberg fields and where
$\eta_{ab}=\mathrm{diag}(-1,1,1,1)$. Further, $S_{\sigma}$ is
defined as
\begin{equation}
S_\sigma=-\frac{\omega}{2}\int d^4x\sqrt{-\hg}\hg^{\mu\nu}\partial_\mu\sigma
\partial_\nu \sigma \ .
\end{equation}
Note that $\Omega(\Phi)$ is  function of $\sigma$ which is necessary
for the invariance of the theory under  global symmetry
\begin{equation}\label{glodil}
\sigma\rightarrow \sigma+\sigma_0 \ , \quad \phi^a\rightarrow
e^{-\sigma_0/M_p}\phi^a
\end{equation}
so that under (\ref{glodil}) $f_{\mu\nu}$ and $\tf_{\mu\nu}$
transform as
\begin{equation}
f_{\mu\nu}\rightarrow e^{-2\sigma_0/M_p}f_{\mu\nu} \ , \quad
\tf_{\mu\nu}\rightarrow e^{-2\sigma_0/M_p}\tf_{\mu\nu} \ .
\end{equation}
The massive term contains square root of the  expression
$\hg^{\mu\nu}\tf_{\nu\rho}$ that under (\ref{glodil}) transforms as
\begin{equation}
\sqrt{\hg^{-1}\tf}\rightarrow e^{-\sigma_0/M_p}\sqrt{\hg^{-1}\tf}
\end{equation}
which implies that $\Omega(\sigma)$ has to have the form
\begin{equation}\label{defOmega}
\Omega(\sigma)=e^{\sigma/M_p} \ .
\end{equation}
Now we are ready to proceed to the Hamiltonian analysis of given
theory. Due to the presence of the square root in the action we
perform the redefinition of the shift functions
\cite{Hassan:2011vm,Hassan:2011hr}
\begin{equation}\label{defNi}
N^i=M\tn^i+\tf^{ik}\tf_{0k}+N \tD^i_{ \ j } \tn^j \ ,
\end{equation}
where
\begin{equation}
 M^2=-\tf_{00}+\tf_{0k}\tf^{kl}\tf_{l0} \ , \quad
 \tf_{ij}\tf^{jk}=\delta_i^{ \ j} \ ,
 \end{equation}
 and where
 $\tD^i_{ \ j}$ obeys the equation
\begin{eqnarray}
\sqrt{\tx}\tD^i_{ \ j}= \sqrt{(g^{ik}-\tD^i_{ \ m} \tn^m \tD^k_{ \
n}\tn^n)\tf_{kj}} \ , \quad
 \tx=1-\tn^i \tf_{ij}\tn^j \
 \nonumber \\
\end{eqnarray}
and also following important identity
\begin{eqnarray}
\tf_{ik}\tD^k_{ \ j}= \tf_{jk}
\tD^k_{ \ i } \ .  \nonumber \\
\end{eqnarray}
We also use $3+1$ decomposition of the four dimensional metric
$\hat{g}_{\mu\nu}$ \cite{Gourgoulhon:2007ue,Arnowitt:1962hi}
\begin{eqnarray}
\hat{g}_{00}=-N^2+N_i g^{ij}N_j \ , \quad \hat{g}_{0i}=N_i \ , \quad
\hat{g}_{ij}=g_{ij} \ ,
\nonumber \\
\hat{g}^{00}=-\frac{1}{N^2} \ , \quad \hat{g}^{0i}=\frac{N^i}{N^2} \
, \quad \hat{g}^{ij}=g^{ij}-\frac{N^i N^j}{N^2} \ .
\nonumber \\
\end{eqnarray}
Note that in $3+1$  formalism the kinetic term for $\sigma$ has the
form
\begin{equation}
-\frac{\omega}{M_p^2} \hg^{\mu\nu}
\partial_\mu\sigma\partial_\nu\sigma=
\frac{\omega}{M_p^2}(\nabla_n\sigma)^2- \frac{\omega}{M_p^2}
\partial_i\sigma g^{ij}\partial_j\sigma \ ,
\end{equation}
where $\nabla_n\sigma$  after redefinition (\ref{defNi}) has the
form
\begin{equation}\label{refnabla}
\nabla_n\sigma=\frac{1}{N} (\partial_t \sigma-
(M\tn^i+\tf^{ik}\tf_{0k}+N \tD^i_{ \ j } \tn^j)\partial_i\sigma) \ .
\end{equation}
With the help of these expressions  we rewrite the action
(\ref{Smass}) into the form
\begin{eqnarray}\label{massgr2}
S&=&M_p^2\int d^3\bx dt \left[N\sqrt{g}
\tK_{ij}\mG^{ijkl}\tK_{kl}+N\sqrt{g}R -\sqrt{g}MU-\right.\nonumber \\
&-&2m^2(N\Omega(\Phi)\sqrt{g}\sqrt{\tx}D^i_{ \ i}-3N\sqrt{g})+\nonumber \\
&+& \left. N\sqrt{g}\frac{\omega}{2M_p^2} (\nabla_n\sigma)^2-
N\sqrt{g}\frac{\omega}{2M^2_p}
\partial_i\sigma g^{ij}\partial_j\sigma\right] \ ,
\nonumber \\
\end{eqnarray}
where
\begin{equation}
U=2m^2\Omega(\Phi)\sqrt{\tx}
 \ ,
 \end{equation}
 and
  where we used the $3+1$ decomposition of the four dimensional
scalar curvature
\begin{equation}\label{31R}
{}^{(4)}R=\tK_{ij}\mG^{ijkl}\tK_{kl}+R \ ,
\end{equation}
where $R$ is three dimensional scalar curvature. We also introduced
de Witt metric
\begin{equation}
\mG^{ijkl}=\frac{1}{2}(g^{ik}g^{jl}+g^{il}g^{jk})- g^{ij}g^{kl}
\end{equation}
with inverse
\begin{equation}
\mG_{ijkl}=\frac{1}{2}(g_{ik}g_{jl}+g_{il}g_{jk})-\frac{1}{2}g_{ij}g_{kl}
\ , \quad
\mG_{ijkl}\mG^{klmn}=\frac{1}{2}(\delta_i^m\delta_j^n+\delta_i^n\delta_j^m)
\ .
\end{equation}
Note that  in (\ref{31R}) we  ignored the terms containing total
derivatives. Finally note that $\tK_{ij}$ is defined as
\begin{equation}
\tK_{ij}=\frac{1}{2N}(\partial_t g_{ij}- \nabla_i
N_j(\tn,g)-\nabla_j N_i(\tn,g)) \ ,
\end{equation}
where $N_i$ depends on $\tn^i$ and $g$ through the relation
(\ref{defNi}).

Now we could proceed to the Hamiltonian formulation of given theory.
However the structure of the derivative $\nabla_n\sigma $
(\ref{refnabla}) suggests  very complicated relations between
momenta and velocities. For that reason we consider simpler case
when we presume that $\sigma$ depends on time only. Note that this
is the reasonable approximation that does not spoil the physical
content of the theory.
From (\ref{massgr2}) we find  the momenta conjugate to $N,\tn^i$ and
$g_{ij}$
\begin{equation}
\pi_N\approx 0 \ , \pi_i\approx 0 \ ,
\pi^{ij}=M_p^2\sqrt{g}\mG^{ijkl}\tK_{kl} \
\end{equation}
while in case of $\phi^a$ and $\sigma$ we find
\begin{eqnarray}
p_a&=&\frac{\mM_{ab}\partial_t\phi^b}{M}
[\tn^i\mR_i+M_p^2 \sqrt{g}U]-f^{ij}\partial_j\phi_a\mR_i \ , \nonumber \\
p_\sigma&=&
-\frac{1}{M}\frac{\alpha_\sigma}{M_p^2 m^2}e^{-2\sigma/M_p^2}
\partial_t\sigma (\tn^i\mR_i+M_p^2\sqrt{g}U)+\omega
\sqrt{g}\partial_t\sigma \ ,
\nonumber \\
\end{eqnarray}
where
\begin{eqnarray}
M^2=M^2_0+\frac{\alpha_\sigma}{M_p^2
m^2}e^{-2\sigma/M_p^2}(\partial_t\sigma)^2 \ , \quad
\mR_i=-2g_{ik}\nabla_j\pi^{kj} \ , \nonumber \\
M_0^2=-\partial_t\phi^a \mM_{ab}\partial_t\phi^b \ , \quad
\mM_{ab}=\eta_{ab}-\partial_i\phi_a f^{ij}\partial_j\phi_b
  \ . \nonumber \\
\end{eqnarray}
These relations imply
\begin{eqnarray}
M_0^2=-\frac{1}{\Pi_a\mM^{ab}\Pi_b+(\tn^i\mR_i+M^2_p\sqrt{g}U)^2}
\frac{\Pi_a\mM^{ab}\Pi_b\alpha_\sigma}
{M_p^2m^2}e^{-2\sigma/M_p^2}(\partial_t\sigma)^2 \nonumber \\
\end{eqnarray}
where $\Pi_a=p_a+f^{ij}\partial_j\phi_a\mR_i$. Then it is easy to
find relation between momenta and velocities
\begin{eqnarray}\label{ptau}
& &p_\sigma +\sqrt{\frac{\alpha_\sigma}{M_p^2m^2}} e^{-\sigma/M_p^2}
\sqrt{\Pi_a\mM^{ab}\Pi_b+(\tn^i\mR_i+M_p^2\sqrt{g}U)^2}=
\frac{1}{N}\omega\sqrt{g}\partial_t\sigma \nonumber \\
& &
\frac{\Pi_a}{\sqrt{\Pi_a\mM^{ab}\Pi_b+(\tn^i\mR_i+M_p^2\sqrt{g}U)^2}}
=\frac{M_pm}{\sqrt{\alpha_\sigma}} e^{\frac{\sigma}{M_p^2}}
\mM_{ab}\frac{\partial_t\phi^b}{\partial_t\sigma} \ .
\nonumber \\
\end{eqnarray}
It is crucial  that these relations do not imply an existence of the
scalar primary constraint which is in sharp contrast with the  case
of the dRGT massive gravity   \cite{Kluson:2012zz} or dRGT massive
gravity coupled to the galileon \cite{Kluson:2013hoa}. On the other
hand,
 using the property of the matrix $\mM_{ab}$ we find three
constraints
\begin{equation}
\partial_i\phi^a\Pi_a\equiv \Sigma_i=\partial_i\phi^a p_a+\mR_i\approx 0
\end{equation}
that, with additional terms proportional to the primary constraints
$\pi_i\approx 0$ are the first class constraints whose smeared forms
are the generator of spatial diffeomorphism.

Now using (\ref{ptau}) we determine   corresponding Hamiltonian
\begin{eqnarray}\label{Hexver}
H=
\int d^3\bx N\mH_0 \ , \nonumber \\
\end{eqnarray}
where
\begin{eqnarray}
\mH_0=\frac{1}{2\omega\sqrt{g}}
\left(p_\sigma+\sqrt{\frac{\alpha_\sigma}{M_p^2m^2}}
e^{-\sigma/M_p^2}
\sqrt{\Pi_a\mM^{ab}\Pi_b+(\tn^i\mR_i+M_p^2\sqrt{g}U)^2}\right)^2+ \nonumber \\
+\frac{1}{\sqrt{g}M_p^2} \pi^{ij}\mG_{ijkl}\pi^{kl}
-\sqrt{g}{}^{(3)}R-2m^2(\Omega\sqrt{g}\sqrt{\tx}\tD^i_{ \
i}-2\sqrt{g})+
\tD^i_{ \ j}\tn^j\mR_i \ . \nonumber \\
\end{eqnarray}
The requirement of the  preservation of the constraint $\pi_N\approx
0$ implies that $\mH_0$ is  constraint as well. The analysis of
constraints is  straightforward. We have six first class constraints
$\Sigma_i\approx 0, \mH_0\approx 0,\pi_N\approx 0$ and six second
class constraints $\pi_i\approx 0,\mC_i\approx 0$ where
$\mC_i\approx 0$ are the secondary constraints that arise from the
requirement of the preservation of the constraints $\pi_i\approx 0$.
These constraints  can be solved for $\pi_i$ and $\tn^i$. Further,
the first class  constraint $\pi_N\approx 0$ can be gauge fixed that
leads to the elimination of $\pi_N$ and $N$ as dynamical variables.
Finally, the constraints $\Sigma_i,\mH_0$ can be again gauge fixed
which leads to the elimination of the St\"{u}ckelberg fields and
conjugate momenta. As a result we are left with $12$ degrees of
freedom coming from the gravity sector that can be identified with
$10$ degrees of freedom corresponding to the massive graviton and
two degrees of freedom corresponding to the scalar at least at the
linearized approximation. Note that this mode cannot be eliminated
due to the absence of the scalar constraint so that Boulware-Deser
ghost is generally present.
 Finally there are two phase space degrees
of freedom $\sigma$ and $p_\sigma$.
\subsection{Note About Gauge Fixing}
After the first version of this paper was published S. Mukohyama
argued in his paper \cite{Mukohyama:2013raa} that the extension of
the quasidilaton theory is ghost free. His arguments is based on the
existence of the additional constraints it his specific gauge
\begin{equation}\label{Mukfixing}
\phi^0=-e^{-\sigma/M_p} \ , \phi^i= \delta^i_\mu x^\mu \ , i=1,2,3 \
.
\end{equation}
Using this gauge he was able to derive the Hamiltonian in the form
$H=\int d^3\bx N\mC_0$ where the specific form of $\mC_0$ is given
in \cite{Mukohyama:2013raa}. According to this result he claims that
there exist an additional constraint in the gauge fixed theory so
that this constraint is responsible for the elimination of the
Boulware-Deser ghost.

In this section we reconsider the gauge fixing (\ref{Mukfixing})
from different point of view. We do not impose the fixing of spatial
diffeomorphism and consider following relation between $\phi_0$ and
$\sigma$
\begin{equation}\label{sigmaphi0}
\sigma=-M_p\ln \phi^0 \
\end{equation}
that is equivalent to (\ref{Mukfixing}).
 Inserting this relation to the definition of
$\tf_{\mu\nu}$ we obtain
\begin{equation}\label{tf}
\tf_{\mu\nu}= \eta_{ab}\partial_\mu\phi^a\partial_\nu\phi^b-
\frac{\alpha_\sigma}{m_g^2}\partial_\mu\phi^0\partial_\nu\phi^0 \ .
\end{equation}
Then inserting  (\ref{tf}) into the dRGT massive gravity we obtain
\begin{equation}\label{Shf}
S=M_p^2 \int d^4x\sqrt{-\hg}
[{}^{(4)}R+2m^2(3-\frac{1}{(\phi^0)^2}\sqrt{\hg^{-1}\tf})-\frac{\omega}{2(\phi^0)^2}
\hg^{\mu\nu}\partial_\mu\phi^0\partial_\nu\phi^0] \ . \nonumber \\
\end{equation}
Now we see that given action is manifestly diffeomorphism invariant.
In other words while we reduce the number of degrees of freedom by
imposing  (\ref{sigmaphi0}) the number of gauge symmetries is the
same. However we mean that the fact that we have less degrees of
freedom than the original theory while the number of gauge
symmetries is the same
 implies that these two theories have different physical content
and should not be considered as equivalent.

From given analysis it is clear that the constraint $\mC_0$ that was
identified in  \cite{Mukohyama:2013raa} corresponds to the
Hamiltonian constraint in the theory with fixed spatial
diffeomorphism. Clearly this constraint should have vanishing
Poisson bracket $\pb{\mC_0(\bx),\mC_0(\by)}$ which implies that it
is the first class constraint. Clearly there is no way how to
generate the additional constraint by imposing the requirement of
the preservation of the constraint $\mC_0$ during the time evolution
of the system since the Hamiltonian is equal to $H=\int d^3\bx
N\mC_0$. However since $\mC_0$ is the first class constraint it can
be gauge fixed and hence we reduce the number of physical degrees of
freedom by two. But we should again stress that the theory
represented by the action (\ref{Shf}) is not equivalent to the
action  (\ref{Smass}) that represents the extension of the
quasidilaton dRGT theory. In summary, we mean that the arguments
that were presented in \cite{Mukohyama:2013raa} do not prove that
the extension of quasidilaton massive gravity is ghost free.

Let us compare this situation with the imposing the static gauge for
the Hamiltonian (\ref{Hexver}). This gauge fixing is represented by
imposing four gauge fixed functions
\begin{equation}\label{gff}
\mG_0=\phi^0-t\approx 0 \ , \mG_i=\phi^i-x^i\approx 0  \ .
\end{equation}
These constraints together with $\mH_0,\tSigma_i$ form the second
class constraints that can be explicitly solved for $p_a$. We solve
$\mH_0$ for $p_0$ since we can identify the gauge fixed Hamiltonian
with $p_0=-\mH_{g.f.}$. The resulting Hamiltonian describes the
dynamic of the physical degrees of freedom $g_{ij},\pi^{ij}$ and
$\sigma,p_\sigma$. The detailed counting of the physical degrees of
freedom was presented in the end of previous section and we will not
repeat it here.

\section{The case $\alpha_\sigma=0$}\label{third}
In previous section we saw that the extension of the quasidilaton
theory that was suggested in \cite{DeFelice:2013tsa} suffers from
the presence of Boulware-Deser ghost due to the absence of two
additional scalar constraints in the Hamiltonian formulation of
given theory. It is instructive to see whether these constraints
emerge in the case of  quasidilaton massive gravity   where
$\alpha_\sigma=0$. Explicitly, we consider the action
\cite{DeFelice:2013awa,D'Amico:2012zv}
\begin{eqnarray}\label{massgr2alpha}
S&=&M_p^2\int d^3\bx dt \left[N\sqrt{g}
\tK_{ij}\mG^{ijkl}\tK_{kl}+N\sqrt{g}R -\sqrt{g}M_0U
-2m^2(N\sqrt{g}\sqrt{\tx}\Omega D^i_{ \ i}-3N\sqrt{g})+\right.\nonumber \\
&+& \left. N\sqrt{g}\frac{\omega}{M_p^2} (\nabla_n\sigma)^2-
N\sqrt{g}\frac{\omega}{M^2_p}
\partial_i\sigma g^{ij}\partial_j\sigma\right] \ .
\nonumber \\
\end{eqnarray}
Let us now perform the Hamiltonian analysis of the action
(\ref{massgr2alpha}). First of all we  find the canonical momenta
\begin{eqnarray}
p_a
&=&-\left(-\mM_{ab}\frac{1}{M_0}\partial_t\phi^b
\tn^i+f^{ij}\partial_j\phi_a
\right)\mR_i+\frac{1}{M_0}M_p^2\sqrt{g}\mM_{ab}\partial_t\phi^b
U+\nonumber \\
&+&\frac{1}{M_0}\mM_{ab}\partial_t\phi^b \tn^i\partial_i\sigma
p_\sigma -\partial_i\sigma f^{ik}\partial_k\phi_a p_\sigma  \ ,
\nonumber \\
p_\sigma&=&\omega\sqrt{g}\nabla_n\sigma \nonumber \\
\end{eqnarray}
so that it is easy to find the scalar  primary constraint
\begin{eqnarray}\label{Sigmapalpha}
& &\Sigma_p\equiv (p_a+(\mR_i+\partial_i\sigma
p_\sigma)f^{ik}\partial_k\phi_a) \eta^{ab}
(p_b+(\mR_i+\partial_i\sigma
p_\sigma)f^{ik}\partial_k\phi_b)+\nonumber \\
&+& (\tn^i(\mR_i+p_\sigma
\partial_i\sigma)+ M_p^2\sqrt{g}U)^2\ \approx 0
\nonumber \\
\end{eqnarray}
and also using the fact that $\partial_i\phi^a \mM_{ab}=0$ we find
additional three constraints
\begin{equation}
\Sigma_i=p_a\partial_i\phi^a+\mR_i+\partial_i\sigma p_\sigma \ .
\end{equation}
Now using these constraints we can simplify (\ref{Sigmapalpha}) so
that it has the form
\begin{equation}
\Sigma_p=p_a\mM^{ab}p_b+(\tn^i(\mR_i+p_\sigma
\partial_i\sigma)+ M_p^2\sqrt{g}U)^2\approx 0 \ .
\end{equation}
This has exactly the same form as the scalar  constraint that
emerges in case of   dRGT massive gravity written in the
St\"{u}ckelberg formalism. The minimal form of this gravity was
analyzed in \cite{Kluson:2012zz} and this analysis can be easily
applied to our case. From (\ref{massgr2}) we
 find the  Hamiltonian with all primary constraints
included
\begin{equation}
H_E=\int d^3\bx (N\mC_0+v_N\pi_N+v^i\pi_i+\Omega_p\Sigma_p+
\Omega^i\tSigma_i) \ ,
\end{equation}
where we introduced the constraint
\begin{equation}
\tSigma_i=\Sigma_i+\partial_i\tn^j\pi_j+
\partial_j(\tn^j\pi_i) \
\end{equation}
and where
\begin{eqnarray}
\mC_0&=& \frac{1}{\sqrt{g}M_p^2} \pi^{ij}\mG_{ijkl}\pi^{kl}-M_p^2
\sqrt{g} R+ 2m^2M_p^2\sqrt{g}\sqrt{\tx}\Omega\tD^i_{ \ i}
-6m^2M_p^2\sqrt{g}+ \nonumber \\
&+&\tD^i_{ \ j}\tn^j(\mR_i+p_\sigma \partial_i\sigma)
+\frac{1}{\sqrt{g}\omega}p_\sigma^2+
\omega\sqrt{g}g^{ij}\partial_i\sigma\partial_j\sigma \ .
\nonumber \\
\end{eqnarray}
As the next step we have to perform the analysis of the stability of
the primary constraints $\pi_i\approx 0,\pi_N\approx 0$ and
$\Sigma_p\approx 0$. In case of the constraint $\pi_i\approx 0$ we
find
\begin{eqnarray}
\partial_t\pi_i=\pb{\pi_i,H}&=&-
\left[(\mR_j+p_\sigma\partial_j\sigma) -\frac{2m^2 M_p^2\Omega
\sqrt{g}}{\sqrt{\tx}}n^k f_{kj}\right]\times \nonumber \\
&\times & \left[N\frac{\delta (\tD^j_{\ m}\tn^m)} {\delta \tn^i}+
\delta^j_i (\tn^i(\mR_i+p_\sigma
\partial_i\sigma)+ M_p^2\sqrt{g}U)\right]=0
\nonumber \\
\end{eqnarray}
 so that we impose following secondary constraint
\begin{equation}
\mC_i=\mR_i+p_\sigma\partial_i\sigma -\frac{2m^2 M_p^2
\Omega\sqrt{g}}{\sqrt{\tx}} f_{ij}\tn^j \ .
\end{equation}
Now with the help of this constraint and the constraint $\tSigma_i$
we can simplify $\Sigma_p$ in the similar  way as in
\cite{Kluson:2012zz}
\begin{eqnarray}
\Sigma_p=4m^4M_p^4g\Omega^2+p_A\eta^{AB}p_B \ .
\end{eqnarray}
Then it is easy to show that $\pb{\Sigma_p(\bx),\Sigma_p(\by)}=0$
and the requirement of the preservation of given constraint leads to
the emergence of an additional constraint. These constraints are
responsible for the elimination of Boulware-Deser ghost,  see again
\cite{Kluson:2012zz} for more details. In other words, the presence
of the kinetic term for the quasidilaton that minimally couples to
gravity does not spoil the  property that given theory is ghost
free.
\section{Ghost Free Extension of Quasidilaton Massive
Gravity}\label{fourth} We argued in section (\ref{second}) that the
extension of quasidilaton theory as was formulated in
\cite{DeFelice:2013tsa} is plagued by the presence of the
Boulware-Deser ghost. On the other hand  given theory has many nice
properties so that it is desirable to propose its  ghost free
version. In this section we propose such a formulation when we
replace the kinetic term for $\sigma$ by following tadpole galileon
term
\begin{eqnarray}\label{Sgall}
S_\sigma &=&-T\int d^4x \Psi(\Phi^A)\sqrt{-\det \tf_{\mu\nu}}
=-T\int d^4x \Psi(\Phi^A)M\sqrt{\tf} \ ,  \nonumber \\
\end{eqnarray}
where $\Phi^A=(\phi^a,\sigma),\tf=\det \tf_{ij}$ and where the
function $\Psi(\Phi^A)$ was chosen in such a way that the action
(\ref{Sgall}) is invariant under (\ref{glodil}). We claim that the
quasidilaton theory formulated as the dRGT massive theory with
$\tf_{\mu\nu}$ and with the kinetic term for the galileon given by
(\ref{Sgall}) is ghost free.

To see this explicitly it is useful to introduce following notation.
Let s write  $\tf_{\mu\nu}$ as
\begin{equation}
\tf_{\mu\nu}=\partial_\mu\Phi^A \mG_{AB}\partial_\nu \Phi^B \  ,
\end{equation}
 where we introduced the metric $\mG_{AB}$
\begin{equation}
\mG_{AB}=\left(\begin{array}{cc} \eta_{AB} & 0 \\
0 & -\frac{\alpha_\sigma}{M_p^2m^2}e^{-2\sigma/M_p} \\ \end{array}
\right) \ .
\end{equation}
We see that our proposal  has the form of the galileon coupled to
dRGT  massive gravity \cite{Andrews:2013ora} whose Hamiltonian
analysis was performed in \cite{Kluson:2013hoa}. On the other hand
the action defined by (\ref{Sgall}) is more complicated since  the
metric $\eta_{AB}$ is replaced with the more general metric
$\mG_{AB}$ that depends on $\Phi^A$ and there are also additional
scalar fields $\Omega(\phi^A), \Psi(\Phi^A)$. However  we can expect
that this fact will not modify the constraint structure of given
theory.

To see this explicitly let us briefly review the Hamiltonian
analysis of the non-linear massive gravity with the term
(\ref{Sgall}) keeping in mind that more detailed analysis can be
found in \cite{Kluson:2013hoa}. As usual  the momenta
 conjugate to $N,\tn^i$ and $g_{ij}$ are
\begin{equation}
\pi_N\approx 0 \ , \pi_i\approx 0 \ ,
\pi^{ij}=M_p^2\sqrt{g}\mG^{ijkl}\tK_{kl} \
\end{equation}
while the  momentum conjugate to $\Phi^A$ has the form
\begin{eqnarray}
p_A
=- \left(\frac{\delta M}{\delta\partial_t \Phi^A}
\tn^i+\mG_{AB}\tf^{ij}\partial_j\Phi^B\right) \mR_i-M_p^2\sqrt{g}
\frac{\delta M}{\partial_t \Phi^A}
U' \ ,  \nonumber \\
\end{eqnarray}
where
\begin{equation}
U'=2m^2\Omega(\Phi)\sqrt{\tx}+\frac{T}{M_p^2} \Psi(\Phi)\sqrt{\tf} \
,
\end{equation}
and where $M^2$ has  the form
\begin{eqnarray}
M^2=-\partial_t\Phi^A \mM_{AB}\partial_t \Phi^B \ , \quad
\mM_{AB}=\mG_{AB}-\mG_{AC}\partial_i\Phi^C
\tf^{ij}\partial_j\Phi^D\mG_{DB} \ .
\nonumber \\
\end{eqnarray}
Note that   the matrix $\mM_{AB}$ obeys following relations
\begin{eqnarray}\label{mMprop}
\mM_{AB}\mG^{BC}\mM_{CD}=
\mM_{AD} \ , \quad \partial_i\Phi^A \mM_{AB}=0 \ .
\  \nonumber \\
\end{eqnarray}
Then it is easy to determine following primary constraints
\begin{equation}\label{defSigmap}
 \Sigma_p=(\tn^i
\mR_i+M_p^2\sqrt{g}U')^2+(p_A+\mR_i\tf^{ij}\mG_{AC}\partial_j\Phi^C)
\mG^{AB}(p_B+\mR_i \tf^{ij}\mG_{BD}\partial_j\Phi^D)\approx
0 \  \nonumber \\
\end{equation}
and
\begin{eqnarray}\label{defSigmai}
\partial_i\Phi^A
p_A+\mR_i=\Sigma_i\approx  0 \ .  \nonumber
\\
\end{eqnarray}
%
Now we are ready to write the extended Hamiltonian which includes
all the primary constraints
\begin{equation}
H_E=\int d^3\bx (N\mC_0+v_N\pi_N+v^i\pi_i+\Omega_p\Sigma_p+
\Omega^i\tSigma_i) \ ,
\end{equation}
where
\begin{eqnarray}
\mC_0= \frac{1}{\sqrt{g}M_p^2} \pi^{ij}\mG_{ijkl}\pi^{kl}-M_p^2
\sqrt{g} R+ 2m^2M_p^2\sqrt{g}\Omega(\Phi)\sqrt{\tx}\tD^i_{ \ i}
-6m^2M_p^2\sqrt{g}+
\tD^i_{ \ j}\tn^j\mR_i \nonumber \\
\end{eqnarray}
and where we introduced the constraints $\tSigma_i$ defined as
\begin{equation}
\tSigma_i=\Sigma_i+\partial_i \tn^i\pi_i+
\partial_j(\tn^j\pi_i) \ .
\end{equation}

To proceed further we have to check the stability of all
constraints. The procedure is the same as in \cite{Kluson:2013hoa}
so that we find that $\tSigma_i$ are the first class constraints
while the requirement of the preservation of the constraints
$\pi_i\approx 0$ implies following secondary constraints
\cite{Hassan:2011vm,Hassan:2011hr}
\begin{equation}\label{defmCi}
\mC_i\equiv \mR_i-\frac{2m^2M_p^2\Omega(\Phi)\sqrt{g}}{\sqrt{\tx}}
\tf_{ij} \tn^j \approx 0 \ .
\end{equation}
Further,  the
 requirement of the preservation of the
constraint $\pi_N\approx 0$ implies an existence of the secondary
constraint $\mC_0\approx 0$.
Using the constraints $\mC_i$ and $\Sigma_i$ we replace the
constraint $\Sigma_p$ by new independent constraint $\tSigma_p$
\begin{equation}\label{Sigmapfin}
\tSigma_p=4m^4M_p^4\Omega^2g+p_A\mG^{AB}p_B+
2T\Psi\sqrt{\tf}\sqrt{p_A\partial_i\phi^A
\tf^{ij}\partial_j\phi^Bp_B+4m^4M_p^4\Omega^2g}+T^2\Psi^2 \tf=0 \ .
\end{equation}
 Then the total Hamiltonian, where we include
all constraints, takes the form
\begin{equation}\label{HTfin}
H_T=\int d^3\bx (N\mC_0+v_N\pi_N+v^i\pi_i+ \Omega_p \tSigma_p+
\Omega^i\tSigma_i+\Gamma^i\mC_i) \ .
\end{equation}
 Now we are ready to
analyze the stability of all constraints that appear in
(\ref{HTfin}). Again, the analysis is the same as in
\cite{Kluson:2013hoa} with slight complication that now there are
additional terms $\Psi(\Phi),\Omega(\Phi)$ together with
$\mG_{AB}(\Phi^A)$ in the definition of the action. However these
terms  are local functions of $\Phi^A$ so that they do not affect
the result that $\pb{\tSigma_p(\bx), \tSigma_p(\by)}\approx 0$. As a
result the requirement of the preservation of the constraint
$\tSigma_p\approx 0$ implies new constraint $\tSigma^{II}_p\approx
0$. These two constraints are the second class constraints that can
be used for the elimination of the Boulware-Deser ghost mode and its
conjugate momenta.

In this section we proposed an extension of the quasidilaton massive
gravity that is ghost free. This proposal  can be generalized in
different ways, either consider the most general potential of the
dRGT massive gravity or more complicated kinetic term for $\sigma$.
It would be also very interesting to analyze the cosmological
consequences of the model with the action (\ref{Sgall}) following
\cite{DeFelice:2013tsa}.
\\
 \noindent {\bf
Acknowledgement:}

 This work   was
supported by the Grant agency of the Czech republic under the grant
P201/12/G028. \vskip 5mm



\begin{thebibliography}{20}



\bibitem{deRham:2010ik}
  C.~de Rham, G.~Gabadadze,
\emph{``Generalization of the
Fierz-Pauli Action,''}
  Phys.\ Rev.\  {\bf D82 } (2010)  044020.
  [arXiv:1007.0443 [hep-th]].


\bibitem{deRham:2010kj}
  C.~de Rham, G.~Gabadadze, A.~J.~Tolley,
\emph{``Resummation of Massive
Gravity,''}
  Phys.\ Rev.\ Lett.\  {\bf 106}, 231101 (2011).
  [arXiv:1011.1232 [hep-th]].



\bibitem{Boulware:1973my}
  D.~G.~Boulware and S.~Deser,
\emph{``Can gravitation have a finite range?,''}
  Phys.\ Rev.\ D {\bf 6} (1972) 3368.




\bibitem{Boulware:1972zf}
  D.~G.~Boulware and S.~Deser,
\emph{``Inconsistency of finite range gravitation,''}
  Phys.\ Lett.\ B {\bf 40} (1972) 227.


\bibitem{DeFelice:2012mx}
  A.~De Felice, A.~E.~Gumrukcuoglu and S.~Mukohyama,
\emph{``Massive gravity: nonlinear instability of the homogeneous
and isotropic universe,''}
  Phys.\ Rev.\ Lett.\  {\bf 109} (2012) 171101
  [arXiv:1206.2080 [hep-th]].

\bibitem{D'Amico:2011jj}
  G.~D'Amico, C.~de Rham, S.~Dubovsky, G.~Gabadadze, D.~Pirtskhalava and A.~J.~Tolley,
\emph{``Massive Cosmologies,''}
  Phys.\ Rev.\ D {\bf 84} (2011) 124046
  [arXiv:1108.5231 [hep-th]].



\bibitem{Gumrukcuoglu:2012aa}
  A.~E.~Gumrukcuoglu, C.~Lin and S.~Mukohyama,
\emph{``Anisotropic Friedmann-Robertson-Walker universe from
nonlinear massive gravity,''}
  Phys.\ Lett.\ B {\bf 717} (2012) 295
  [arXiv:1206.2723 [hep-th]].

\bibitem{DeFelice:2013awa}
  A.~De Felice, A.~E.~Gümrükçüog(lu, C.~Lin and S.~Mukohyama,
\emph{``Nonlinear stability of cosmological solutions in massive
gravity,''}
  JCAP {\bf 1305} (2013) 035
  [arXiv:1303.4154 [hep-th]].

\bibitem{D'Amico:2012zv}
  G.~D'Amico, G.~Gabadadze, L.~Hui and D.~Pirtskhalava,
\emph{``Quasidilaton: Theory and cosmology,''}
  Phys.\ Rev.\ D {\bf 87} (2013) 064037
  [arXiv:1206.4253 [hep-th]].

\bibitem{Huang:2012pe}
  Q.~-G.~Huang, Y.~-S.~Piao and S.~-Y.~Zhou,
\emph{``Mass-Varying Massive Gravity,''}
  Phys.\ Rev.\ D {\bf 86} (2012) 124014
  [arXiv:1206.5678 [hep-th]].

\bibitem{DeFelice:2013tsa}
  A.~De Felice and S.~Mukohyama,
\emph{``Towards consistent extension of quasidilaton massive
gravity,''}
  arXiv:1306.5502 [hep-th].

\bibitem{Gabadadze:2012tr}
  G.~Gabadadze, K.~Hinterbichler, J.~Khoury, D.~Pirtskhalava and M.~Trodden,
\emph{``A Covariant Master Theory for Novel Galilean Invariant
Models and Massive Gravity,''}
  Phys.\ Rev.\ D {\bf 86} (2012) 124004
  [arXiv:1208.5773 [hep-th]].




\bibitem{Kluson:2012zz}
  J.~Kluson,
\emph{``Note About Hamiltonian Formalism for General Non-Linear
Massive Gravity Action in Stuckelberg Formalism,''}
  arXiv:1209.3612 [hep-th].



\bibitem{Kluson:2012wf}
  J.~Kluson,
\emph{``Non-Linear Massive Gravity with Additional Primary
Constraint and Absence of Ghosts,''}
  Phys.\ Rev.\ D {\bf 86} (2012) 044024
  [arXiv:1204.2957 [hep-th]].

\bibitem{Hassan:2012qv}
  S.~F.~Hassan, A.~Schmidt-May and M.~von Strauss,
\emph{``Proof of Consistency of Nonlinear Massive Gravity in the
St\'uckelberg Formulation,''}
  Phys.\ Lett.\ B {\bf 715} (2012) 335
  [arXiv:1203.5283 [hep-th]].




\bibitem{Kluson:2013hoa}
  J.~Kluson,
\emph{``Hamiltonian Analysis of Minimal Massive Gravity Coupled to
Galileon Tadpole Term,''}
  arXiv:1305.6751 [hep-th].

\bibitem{Andrews:2013ora}
  M.~Andrews, G.~Goon, K.~Hinterbichler, J.~Stokes and M.~Trodden,
\emph{``Massive gravity coupled to DBI Galileons is ghost free,''}
  Phys.\ Rev.\ Lett.\  {\bf 111} (2013) 061107
  [arXiv:1303.1177 [hep-th]].

\bibitem{Gourgoulhon:2007ue}
  E.~Gourgoulhon,
\emph{``3+1 formalism and bases of numerical relativity,''}

  [gr-qc/0703035 [GR-QC]].

\bibitem{Arnowitt:1962hi}
  R.~L.~Arnowitt, S.~Deser, C.~W.~Misner,
 \emph{``The Dynamics of general
 relativity,''}
  [gr-qc/0405109].






\bibitem{Hassan:2011vm}
  S.~F.~Hassan, R.~A.~Rosen,
\emph{``On Non-Linear Actions for Massive Gravity,''}
  JHEP {\bf 1107 } (2011)  009.
  [arXiv:1103.6055 [hep-th]].


\bibitem{Hassan:2011hr}
  S.~F.~Hassan, R.~A.~Rosen,
\emph{``Resolving the Ghost Problem in
non-Linear Massive Gravity,''} Phys.\
Rev.\ Lett.\  {\bf 108} (2012) 041101
  [arXiv:1106.3344 [hep-th]].

\bibitem{Huang:2013mha}
  Q.~-G.~Huang, K.~-C.~Zhang and S.~-Y.~Zhou,
\emph{``Generalized massive gravity in arbitrary dimensions and its
Hamiltonian formulation,''}
  JCAP {\bf 1308} (2013) 050
  [arXiv:1306.4740 [hep-th]].



\bibitem{Koyama:2011wx}
  K.~Koyama, G.~Niz and G.~Tasinato,
\emph{``The Self-Accelerating Universe with Vectors in Massive
Gravity,''}
  JHEP {\bf 1112} (2011) 065
  [arXiv:1110.2618 [hep-th]].

\bibitem{Tasinato:2012ze}
  G.~Tasinato, K.~Koyama and G.~Niz,
\emph{``Vector instabilities and self-acceleration in the decoupling
limit of massive gravity,''}
  Phys.\ Rev.\ D {\bf 87} (2013) 064029
  [arXiv:1210.3627 [hep-th]].

\bibitem{Khosravi:2013axa}
  N.~Khosravi, G.~Niz, K.~Koyama and G.~Tasinato,
\emph{``Stability of the Self-accelerating Universe in Massive
Gravity,''}
  JCAP {\bf 1308} (2013) 044
  [arXiv:1305.4950 [hep-th]].

\bibitem{Mukohyama:2013raa}
  S.~Mukohyama,
\emph{``Extended quasidilaton massive gravity is ghost free,''}
  arXiv:1309.2146 [hep-th].













%
%
%
%
%
%
%
%
%
%
%
%
%
%
%
%
%
%
%
%
%
%
%
%
%
%
%
%
%
%
%
%
%
%
%
%
%
%
%
%
%
%
%
%
%
%
%
%
%
%
%
%
%
%
%
%
%
%
%
%
%
%
%
%
%
%
%
%
%
%
%
%
%
%
%
%
%
%
%
%
%
%
%
%
%
%
%
%
%
%
%
%
%
%
%
%
%
%
%
%
%
%
%
%
%
%
%
%
%
%
%
%
%
%
%
%
%
%
%
%
%
%
%
%
%
%
%
%
%
%
%
%
%
%
%
%
%
%
%
%
%

%
%
%
%
%
%
%
%
%
%
%
%
%
%
%
\end{thebibliography}
\end{document}